\documentclass[namedreferences,hyperref,optionalrh]{spr-sola}
\usepackage{graphicx}        
\usepackage{color}           

\usepackage{xspace}           

\chardef\us=`\_

\newcommand{\arcsec}{$^{\prime \prime}$ \xspace}

\usepackage{hyperref}

\begin{document}

\begin{frontmatter}
\title{Small EUV Brightenings in the Quiet Solar Atmosphere: New Insights from the Solar Orbiter Mission}

\author[addressref={aff1},  email={susanna.parenti@universite-paris-saclay.fr}]{\inits{S.}\fnm{Susanna}~\snm{Parenti}\orcid{0000-0003-1438-1310}}
\address[id=aff1]{Universit\'e Paris-Saclay, CNRS, Institut d’Astrophysique Spatiale, 91405 Orsay, France}

\runningauthor{Parenti S.}
\runningtitle{\textit{Small EUV Brightenings in the Quiet Sun} }

\begin{abstract}

One of the many outcomes of the  Solar Orbiter mission is the evidence for the solar atmosphere being filled by highly impulsive bursts, down to $\approx$ 200 $\mathrm{km}$ scale: the limit of the EUV instruments' spatial resolution. Small-scale events of this kind were already known, but their  observation was occasional or with limited, lower resolution.  Solar Orbiter has revealed that small scale, highly impulsive events are everywhere on  the quiet Sun, all the time, at even smaller scales. Their similarity with known larger features, are the witnesses that the physical processes causing them are independent of the spatial scales involved.
 Their  highly dynamic property is the signature of energy transfer and/or local dissipation. Their investigation can thus elucidate on the dominant physical processes acting on the solar atmosphere and on the possible role in the origin of the hot solar corona.

In this review, we will present a summary of the observational and  simulation results on this topic, led by the results from data taken by the The Extreme Ultraviolet Imager
 (EUI)/High Resolution Imagers (HRI\textsubscript {EUV}) instrument. Here, we will cover both statistical properties and analyses of individual events. 

\end{abstract}

\end{frontmatter}

\section{Introduction}
     \label{intro}

Decades of observations of the solar atmosphere have shown us that small EUV brightenings, transient increases of the brighteness, are common. These have been classified based on several parameters, sometimes linked to their location,  shape, presence of plasma flows, and others. They are found throughout the whole solar atmosphere, sometimes localized in the cooler lower atmosphere or in the hot corona, sometimes having observational signatures in all these layers, including possible emission in X-rays. Some recent reviews on QS brightening have been proposed by \cite{Young2018}, \cite{Nelson22}, and \cite{Harra_2025}.

\begin{figure*}
    \centering
    \includegraphics[width=0.5\textwidth]{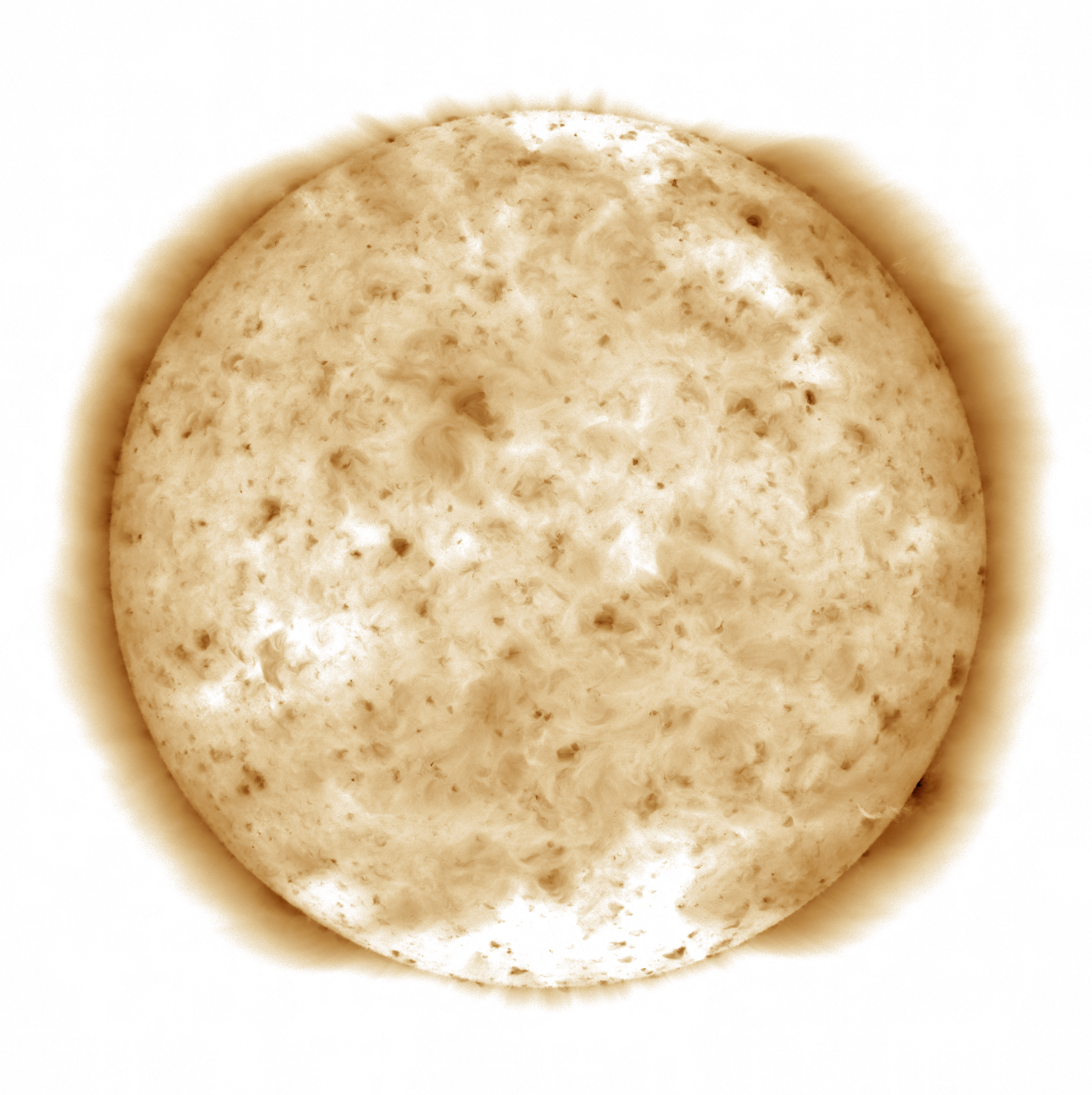}
    \caption{ \bf The solar corona during the minimum of the solar cycle on February 10, 2020, observed by  SDO/AIA in the 193 \AA~ band (inverse colors). The QS dominates the surface, while the faint coronal holes are visible at the poles. The bright dots (dark dots in the figure) are the location of bright points mentioned in the text. Courtesy of JHelioviewer.}
\label{fig:QS}   
\end{figure*}

The growing attention to these events in the corona is mostly due to their interpretation as the possible signature of coronal heating: the solar corona is maintained at an average temperature of about 1 MK, due probably to one or more dominant, still partially understood, physical processes. 
This is particularly interesting because even at the minimum of the solar activity, when the coronal surface is composed of the quiet Sun (QS) and cooler coronal hole (CH), this temperature is maintained.   This is visible in Figure \ref{fig:QS} which shows the corona at around 1.5 MK  as seen by the  Solar Dynamic Observatory/Atmospheric Imaging Assembly (SDO/AIA, \citealp{Lemens2012}) in  the band 193 \AA. During this period the only detectable hot ($> 1.5 ~\mathrm{MK}$) emission in the QS arises from {\it coronal bright points} (small versions of active regions, \citealp{Madjaska19}) and their microflares: impulsive brightening, as shown in Figure \ref{fig:tiwari22}. Investigations of such hot emission have revealed that they do not release enough energy to sustain the corona's high temperature (see \citealp{hudson91} and Section \ref{stat} of this review). At the same time, there is observational evidence that the corona is dominated by impulsive transients at all scales, with an energy--frequency distribution following a power\,--law  with a negative slope generally around two. \cite{hudson91} has  demonstrated that, possibly, unresolved events may have a major role in the coronal heating and in the sustain of the high temperature. This argument has been the driver of many observational investigations, spanning  the EUV and X-ray bands, with the aim of pushing the existing instruments to work at their detection limits in the hope of measuring the still unreached lower energy ranges (see, for instance, \cite{hannah10} for the QS). One of the results has been to understand that probably, the coronal features at all observed spatial scales are made of a fine, probably unresolved by the current instrumentation, structure. 

Only the so-called 'diffuse emission' seen in the EUV radiative losses of the QS, can be still considered another component of the QS.  
This is a diffuse, relatively stable emission over spatial scales below that of supergranules ($20~ \mathrm{Mm}$, and temporal scales of at least 25 minutes \citep{Gorman2023, Dolliou23}). The question on what causes such properties is still unanswered, even though \cite{Gorman2023} provides arguments in favor of propagating waves or MHD turbulence against a more intermittent and impulsive phenomena, such as jets or bursts. The question is still open, as the reached spatial resolutions are not enough to claim if this is really a diffuse, unstructured emission, or the result of unresolved fine scale features. This review will not cover this topic.

Among the rich bibliography on the small-scale brightening, it is relevant  to mention the observations by the High-Resolution Coronal Imager (Hi-C) sounding rocket flights \citep{Kobayashi2014}, which imaged the solar corona in the 193 EUV band (centered on  Fe XII)  or 172 EUV band (centered on Fe IX-Fe X) at a spatial resolution of about 0.3\arcsec (about 220 km) for the few minutes of the flight. Observing  active regions, these data revealed the corona to have fine structures at the limit of the instrument's resolution (\citealp{Winebarger2013, Winebarger_2014}), including small loops and bright transient dots (the latter at the loop footpoints, with a size of about 700 km across and 25 seconds duration, \citealp{Regnier2014}). Jetlet-type of features, an apparent smaller version of the   Interface Region Imaging Spectrograph (IRIS: \citealp{DePontieu2014}) jetlets, were observed outside large loops, rooted at the edges of magnetic network lane \citep{Panesar2019}. 
\cite{reale07},  using the  HINODE/XRT multi-band instrument \citep{Golub07}, presented evidence for the existence of this fine  thermal structure in active regions (even though  over a lower available resolution of about 1$^{\prime\prime}$)  also in the soft X-ray band.
 
 Hi-C observed hotter plasma than  IRIS which has similar spatial resolution and which routinely observes of the solar atmosphere. However, this instrument is mostly sensitive to chromospheric and transition\,--\,region emission, and it is mostly blind to what happens in corona. Today, the  EUI/HRI\textsubscript {EUV} \citep{EUI_instrument} instrument on board Solar Orbiter \citep{Muller} can recover some of this missing information.

\begin{figure*}
    \centering
    \includegraphics[width=0.6\textwidth]{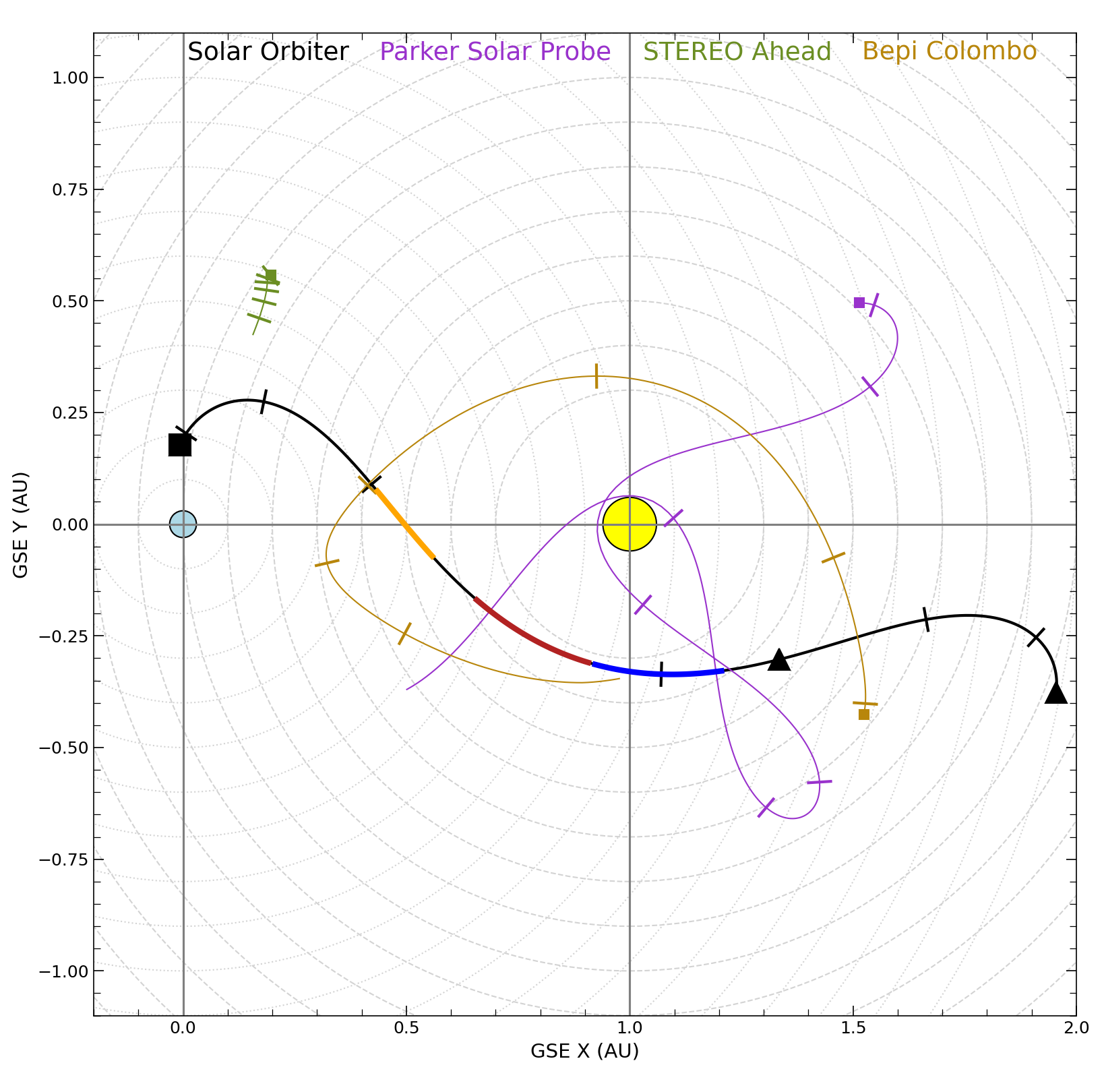}
    \caption{First orbit ({\it dark curve}) of  Solar Orbiter during the nominal mission phase in a geocentric solar ecliptic coordinates (27 December 2021 -- 27 June 2022). The orange, red and blue part of the orbit mark the period when the remote sending instruments observe.
    The other colored curves mark to the orbits of  STEREO Ahead,  Parker Solar Probe and  Bepi Colombo.  The Earth is at 0.0, the Sun at 1.0. Courtesy of ESA (\url{https://www.cosmos.esa.int/web/solar-orbiter}).}
\label{fig:orbit}   
\end{figure*}

The ESA  Solar Orbiter mission, launched in 2020, has unique orbits that approach
the Sun down to about 0.29 AU. The payload, consisting of ten instruments, is collecting data to answer four top science questions using both coordinated observations and a common pointing. Details on the modes of observations are proposed by \cite{Zouganelis2020} and \cite{Auchere2020}. 
Among the onboard instruments, the  EUI imagers suite includes the  Full Sun Imager (FSI) and two high-resolution telescopes one of which, HRI\textsubscript {EUV},  has about 1\arcsec  angular resolution. This corresponds to a different resolution at the Sun, depending on the spacecraft solar distance.
The combination of this spatial resolution and the close approach to the Sun, allows for observing features at very high  spatial resolution in the 174 band (Fe IX-X) reaching that of  Hi-C. Furthermore, it can run at  a cadence $>$ 1 second, thus being able to detect very thin details of dynamical events. 
The regular observations taken with this instrument have revealed a solar corona that is highly dynamic and highly structured (see for instance \citealp{Berghmans_2023}). HRI\textsubscript {EUV}, together with the high resolution Polarimetric and Helioseismic Imager (PHI: \citealp{Solanki20}) and the  Spectral Imaging of the Coronal Environment (SPICE: \citealp{spice}) on board Solar Orbiter, and other Earth view assets, have thus high potential for investigating the small brightenings that 
we are interested in throughout the whole solar atmosphere. The coordinated use of the existing assets is fundamental for collecting as much  information as possible on these features. Figure \ref{fig:orbit} shows an example of the first orbit during the nominal mission phase of Solar Orbiter in 2022. The variation of the line of sight with respect to Earth and distance from the Sun are often used during these coordinated observations. 
 The goal of this review is to provide an overview of the new insights on this topic provided by observing the QS with such  assets.

%
%

\section{EUV Brightening: General Properties}
\label{sec:gen}

\begin{figure*}
    \centering
    \includegraphics[width=\textwidth]{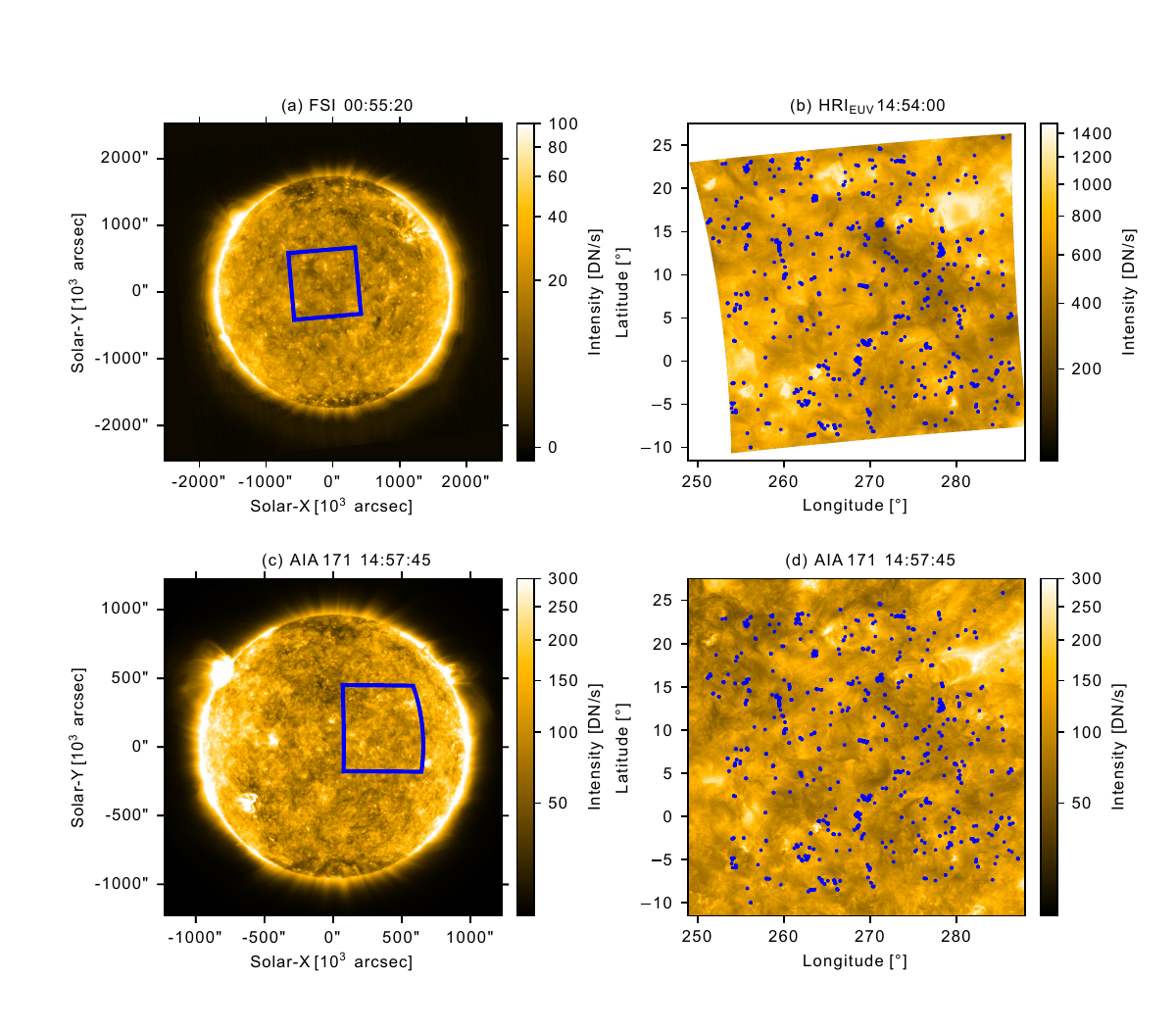}
    \caption{{\it Left}: the Sun observed in the 174~ \AA band of  Solar Orbiter/FSI on 30 May 2020. The {\it blue square} locates the HRI\textsubscript {EUV} field of view, shown in detail on the right panel. Here, the {\it blu dots} represent the brightening detected within the five-minutes sequence. Reproduced with permission from \cite{Dolliou23}, copyright by the author(s).}
\label{fig:Dolliou23}   
\end{figure*}

\begin{figure*}
     \centering
        \includegraphics[width=\textwidth,clip=]{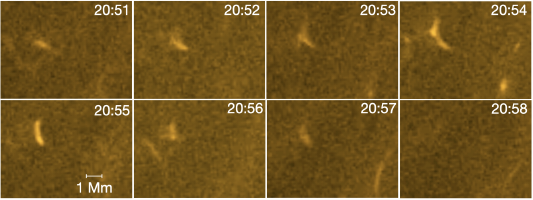}
         \includegraphics[width=\textwidth,clip=]{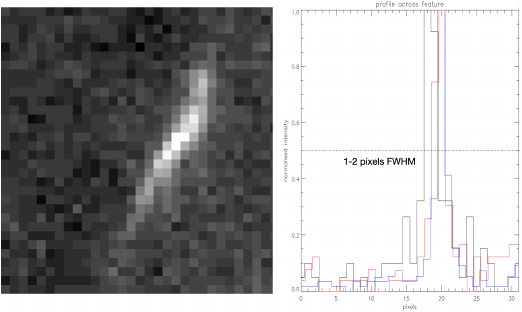}
\caption{{\it Left}: example of HRI\textsubscript {EUV} brightenings observed in a QS area at a distance of 0.556 AU. {\it Right}: calibrated images (L2 level) with the original sensor pixelization of an event observed from 0.323 AU; each pixel corresponds to 115  $\mathrm{km}$ on the Sun. {\it Right}: Various cuts through the middle-panel event, demonstrating that the FWHM of the feature is 1\,--\,2 pixels. Reproduced with permission from \cite{Berghmans_2023}, copyright by the author(s).}
\label{fig:bergh23}
\end{figure*}

One of the key aspects to start with is the definition we give here of small EUV brightenings: an impulsive  local enhancement of brightness in the EUV up to a few megameters in size (see for instance Figure \ref{fig:bergh23} and the following). This is what was set in the first high temporal and spatial resolution observation of the HRI\textsubscript {EUV} by \cite{Berghmans2021}. This article illustrates the general properties of the events in terms of projected area, duration, temperature, and location over a temporal sequence of about five minutes at five seconds cadence. In this sequence, more than 1400 events were automatically detected and tracked over time using an algorithm that allows for separating solar brightenings from random intensity variation due to noise and other instrumental effects. 
The algorithm takes a few input parameters, which, among others, set the maximum spatial scale to be detected, about 4 Mm in this case. The spatial resolution was given by the solar distance of 0.55 AU of the spacecraft, which provided the smallest brightening observed of about 0.4 Mm.  These first observations revealed that projected area, total intensity and duration have a negative power  law distribution with the extremes at the imposed limits of the observations or automatic detection method. Thus, for the projected area the limits of the distribution were 0.08 $\mathrm{Mm^2}$ and 5 $\mathrm{Mm^2}$. For a similar reason,  the lower limit for the duration was given by the 10 seconds temporal resolution, and the upper limit by the sequence duration of about 200 seconds. Another result presented in this paper was a first estimate of the temperature, to be about $\log T=6.1$. As presented in the Section \ref{sec:T}, this latter result has been subject of further analysis indicating also lower values.
As illustrated in Table 1 of \cite{Harra_2025}, the selected upper limits in size  are the lowest ones for the size of blinkers and coronal bright points, while only chromospheric jets and explosive events are of the similar size to the HRI\textsubscript {EUV} smallest brightenings.
Figure \ref{fig:Dolliou23} (left panel) shows the QS target for the high resolution observation and, on the right, the localization of these events within this field of view. As visible in this figure,  some events are also visible in AIA 
images, which are about 1400  $\mathrm{km}$ spatial extent. However, the smallest HRI\textsubscript {EUV} events are not seen in AIA and many are detectable only after they are identified by HRI\textsubscript {EUV}. This is due to the lower spatial resolution and cadence of AIA  (about 1100 $\mathrm{km}$ and 12 seconds) with respect to HRI\textsubscript {EUV} (200 $\mathrm{km}$ at perihelion and 2-3 seconds).

\begin{figure}
    \centering
    \includegraphics[width=0.5\textwidth,clip=]{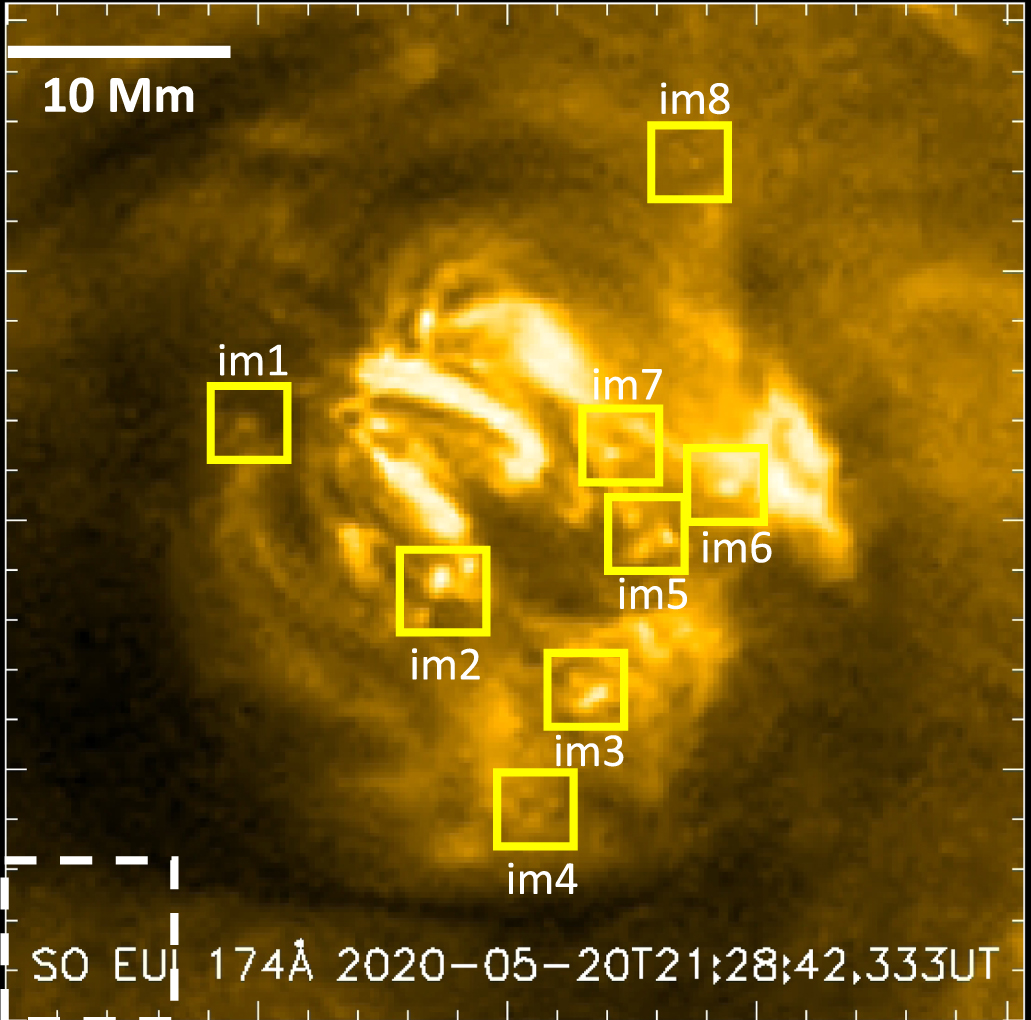}
    \caption{Example of HRI\textsubscript {EUV} brightenings observed in bright-point loops. Reproduced with permission from \cite{Tiwari22}, copyright by the author(s).}
\label{fig:tiwari22}   
\end{figure}

Since then, regular observations have been made to explore their properties in various areas of the Sun, and they are indeed seen everywhere. Many of the results presented in this review come from those observations, where the same detection algorithm developed by Berghmans et al. was applied.  
Figure \ref{fig:bergh23} shows examples of the HRI\textsubscript {EUV} brightenings as seen within the QS while 
Figure \ref{fig:tiwari22} illustrates the presence of such events within bright points. Other examples are found in the figures that follow.  

Given the properties of the detection method, which does not depend on physical processes, the brightenings may have various different properties for their shape, duration, temperature regime, and so on. Examples of such properties can be found, for instance, in Table 1 of \cite{Panesar21}.
This means that they may probably include EUV bursts, jets, and all sorts of small-scale events new or already classified. The capability of the HRI\textsubscript {EUV} instrument to reach unprecedented temporal and spatial resolutions, together with the regularity of observation, allows for more systematic investigations regarding if and how these tiny phenomena can make up the observed corona.


\section{Statistical Properties }
\label{stat}

This section  presents results from statistical studies and their interpretation in terms of contribution to coronal heating. Sections \ref{sec:spa_t}, \ref{sec:pdf} and \ref{sec:cor_h} involve the analysis of the whole  observed brightening population, while Section \ref{sec:selected} provides results from subgroups of such population.

\subsection{General Spatial and Temporal Properties}
\label{sec:spa_t}
 
\begin{figure}
    \centering
    \includegraphics[width=0.5\textwidth]{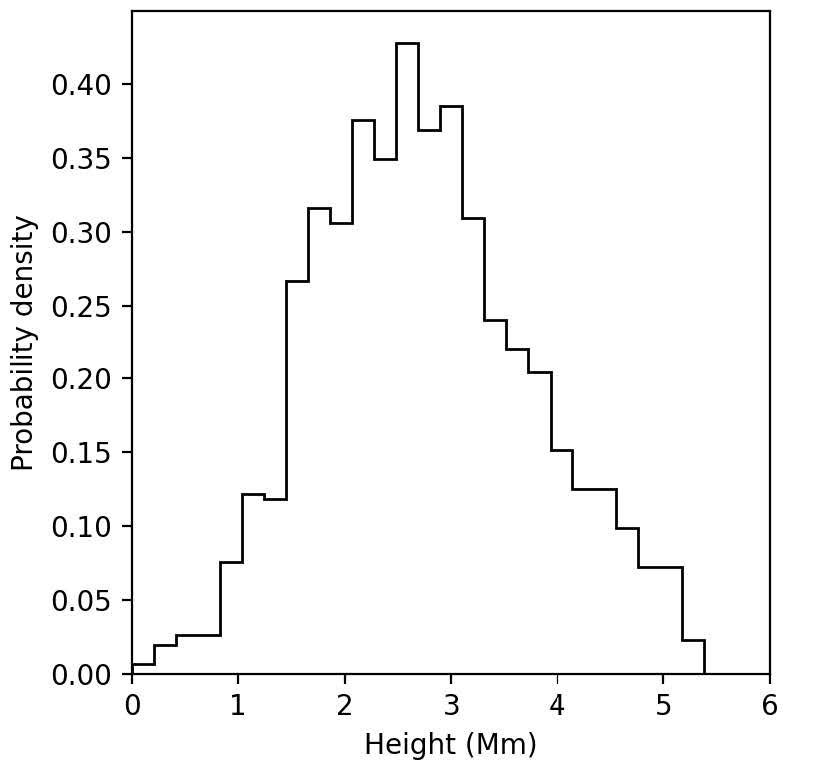}
    \caption{Histogram of the EUV-bringtenings height using  the automatic triangulation between HRI\textsubscript {EUV} and AIA. Reproduced with permission from \cite{Zhukov2021},  copyright by the author(s).}
\label{fig:zhukov21}   
\end{figure}

One of the advantage of the peculiar orbit of Solar Orbiter is the changing Earth-Sun-Spacecraft angle.  This provides the opportunity for stereoscopic measurements.  \cite{Zhukov2021} took this occasion to derive the height above the photosphere of the HRI\textsubscript {EUV} QS brightenings, estimating  a narrow distributed  around a few megameters (1\,--\,5 $\mathrm{Mm}$), as shown in Figure \ref{fig:zhukov21}. This result was achieved by applying the method of triangulation to two datasets of the same area observed from the viewpoints of  Solar Orbiter and SDO/AIA. They were separated by about 31 degrees in longitude. Indeed, the largest events detected by HRI\textsubscript {EUV} are also visible in AIA. To be remarked that the population of these brightenings had all the possible shape and behavior.
Beside, results were discussed in terms of the magnetic shape of the emitting structure, with the more reasonable interpretation being that most of the emission originates from the top of low-lying, small-scale loops. As suggested by these authors, such properties and their appearance resembles the already known chromospheric and transition\,--\,region small-scale features previously observed by, for instance, IRIS \citep{Hansteen14}.

Given the 14$^\prime\times 14^\prime$ field of view (FOV) of HRI\textsubscript {EUV}, there is no yet a large scale statistical investigation on the spatial distribution of these EUV brightenings on whole QS surface at time interval. 
Within the HRI\textsubscript {EUV} FOV, \cite{Alipour22} estimated that only about 27$\%$ of them are found within bright points. However, when we exclude larger scale structures similar to what we see  in Figure \ref{fig:Dolliou23}, these brightenings appear to be localized and they reappear in particular areas identified as the internetwork areas (similarly to \citealp{Berghmans2021}). Often, several subsequent events are observed in the same location as stated by several works reported here (e.g. \cite{Berghmans2021}, \cite{Dolliou24}). 

\begin{figure*}
    \centering
    \includegraphics[width=\textwidth,clip=]{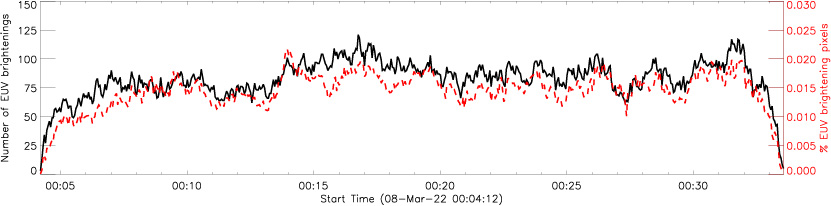}
    \caption{Time series of the number of EUV brightenings ({\it black solid line}) and the percentage of pixels found to be within EUV brightenings ({\it red dashed line}). The decrease of events at the beginning and end of the series is an artifact of the event--selection criteria. Reproduced with permission from \cite{Nelson24a}, copyright by the author(s).}
\label{fig:nelson24a}   
\end{figure*}

\cite{Nelson24a} also counted the number of events in the HRI\textsubscript {EUV} field of view as shown in Figure \ref{fig:nelson24a}. Here we see that the number of brightenings and the area that they occupy remains basically stable over time, but they involve only a small fraction of the total observed surface. Similar results were found by other recent works using EUI/HRI\textsubscript {EUV} and AIA \citep{Joulin2016, Upendran2021, Narang_2025}. 
These statistical works, when applied to the energy inferred by such 
brightenings, provide a quantification of their contribution to coronal heating: at today, the estimate of the energy rate provided by these events has been proven to be insufficient for the maintenance of the corona. 


\subsection{Power-Law  Distributions}
\label{sec:pdf}

  Another approach often used to understand the general behavior of these brightenings and their possible contribution of the corona heating 
is through the use of the Probability Distribution Function (PDM, see Section \ref{intro} and \citealp{Aschwanden2005}). With this method it was discovered that a power\,-- law distribution with negative index (close to 2), dominates both the distributions of several quantities (energy, lifetime, intensity for instance) of coronal EUV images of QS \cite[e.g.][]{Berghmans98, Fludra2023}, and of selected small scale coronal features, over several orders of magnitude. This is called self-similarity behavior.
 At today, the extension over the small scale of the power\,--law  distribution is set by the known instrumental limits (sensitivity, spatial and temporal resolutions, in both X-ray and EUV bands, \citealp[e.g.][]{Hannah_2011, Berghmans98}). This can be affirmed because data from the highest resolution instrument provides an extension to lower values of the same known power\,--law  distribution. If instead, some physical process acts at small scales changing the behavior of the power\,--law  distribution, and these scales are resolved, we should see a change in the index behavior. This is not the case yet, as it will be reported in the following. 

 Among the past interesting results in this field,  it is important to recall those from \cite{Berghmans98} and \cite{Joulin2016} which, using respectively Solar and Heliospheric Observatory/Extreme ultraviolet Imaging Telescope   (SOHO/EIT)  and  SDO/AIA data, have shown the existence of a positive correlation between the event area and duration.  It should be noted that  EIT had a lower cadence (one minute for this work) and spatial resolution  (3.75 $\mathrm{Mm}$) with respect to  AIA (12 \,seconds and about 1 $\mathrm{Mm}$). 
\cite{Joulin2016} have shown the  AIA power\,--law distribution to be a continuation at smaller spatial and temporal scales of those derived from  EIT.  In their datasets,  the smallest events were the shortest, thus highlighting that in order to observe even smaller events, we  need to increase the cadence of the observations. This is what has been achieved with the  Solar Orbiter EUI/HRI\textsubscript {EUV}. 

\begin{figure*}
    \centering
    \includegraphics[width=\textwidth]{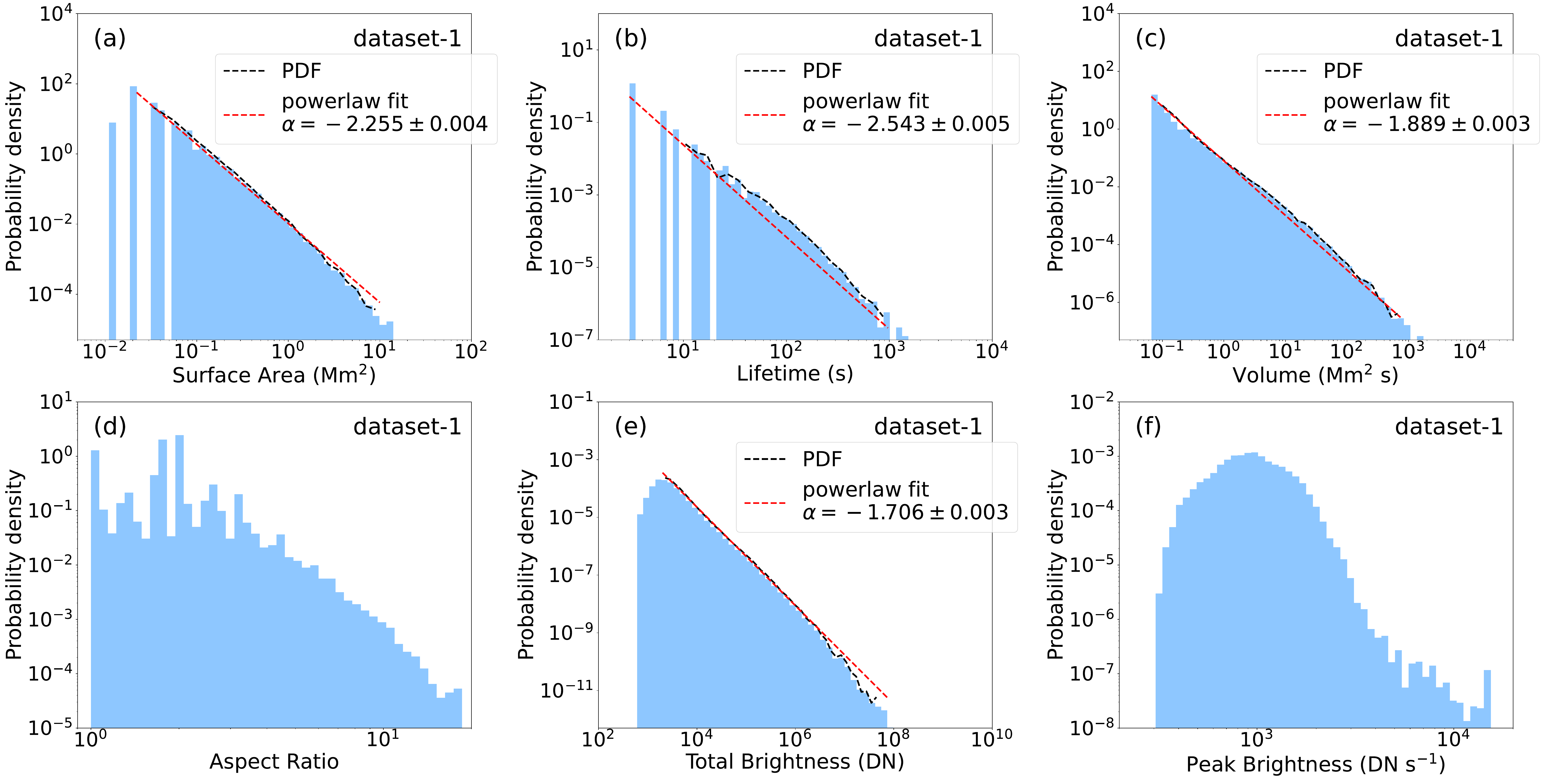}
    \caption{Statistical properties of the HRI\textsubscript {EUV} events as seen at the closest distance to the Sun. Reproduced with permission from \cite{Narang_2025}, copyright by the author(s).}
\label{fig:Narang25}   
\end{figure*}

 Indeed,  \cite{Narang_2025} has  reported the most updated statistical work done over 45 and 60 minutes with HRI\textsubscript {EUV} temporal sequences, at three-second cadence, with about 200 $\mathrm{km}$ spatial resolution (at  Solar Orbiter minimum perihelion) over 79,089 and 98,639 detected events, respectively. The detection algorithm was the same as used by \cite{Berghmans2021}.
Figure \ref{fig:Narang25} shows the probability -- density distributions of the surface area, the lifetime, volume, aspect ratio, total brightness, and peak brighteness of  for the first dataset mentioned. These  distributions confirm and extend to  smaller scales what was found with the previous higher resolution instruments:  surface area, lifetime, volume, total brightness follow a power\,--law  with a negative index close in values to the previous works. 
The smallest and shortest event is about  $0.01 ~\mathrm{Mm^{2}}$ in area and three seconds in duration, even though these quantities are still limited by the instrumental resolution.  Table \ref{tab1}  reports the range where a power\,--law  dominates the probability distributions of several measured quantities. 

 Alike results arise from the statistical work of \cite{Alipour22} who found similar distributions, even though from a smaller dataset (8678 events detected within a length between 400 and 4000 $\mathrm{km}$), adopting a different detection method based on machine learning  \citep{Alipour12}. Besides some small differences in the absolute value of the power\,--law   index that can be found in literature (all around to two), the important result is the presence of this power--law behavior over several orders of magnitude (from flares to picoflares), as a signature of  self-similarity of the solar atmosphere. HRI\textsubscript {EUV} has provided evidence that these already known properties can be extended to smaller spatial and temporal scales. 

\begin{table}[]
    \centering
    \begin{tabular}{c|c|c|c|c}
    \hline
         Surface area$^1$ \tabnotes{1. \cite{Narang_2025}, 2. \cite{Alipour22}}  &  Lifetime$^1$  & Volume$^1$ & Total brightness$^1$ & Peak Intensity$^2$ \\
         ($\mathrm{Mm^2}$) & ($\mathrm{s}$) & ($\mathrm{Mm^2}~s$) & (DN) & (DN) \\
    \hline     
        0.01 -- 5 $\times \mathrm{10^{1}}$ & 3 -- 2 $\mathrm{\times 10^{3}}$ & $\mathrm{10^{-1}}$ -- $\mathrm{10^3}$ & $\mathrm{10^{4}}$ -- $\mathrm{10^{8}}$& 2 $\mathrm{\times 10^{4}}$ -- 2 $\mathrm{\times 10^{5}}$ \\
        \hline    
    \end{tabular}
    \caption{Estimated ranges where the PDF has a power\,--law  distribution for small HRI\textsubscript {EUV} brightenings.}
    \label{tab1}
\end{table}

\subsection{Light Curves and Temperature}
\label{sec:cor_h}



In Section \ref{sec:spa_t} we saw that these HRI\textsubscript {EUV} events alone are not able to sustain the corona. Nevertheless, these  still remain interesting to be studied, as they may partially contribute to build the corona, and they are certainly the signature of a dynamic solar atmosphere that is continuously transferring energy or dissipating it, evolving and reorganizing its magnetic field at those spatial scales.

\begin{figure*}
    \centering
    \includegraphics[width=\textwidth,clip=]{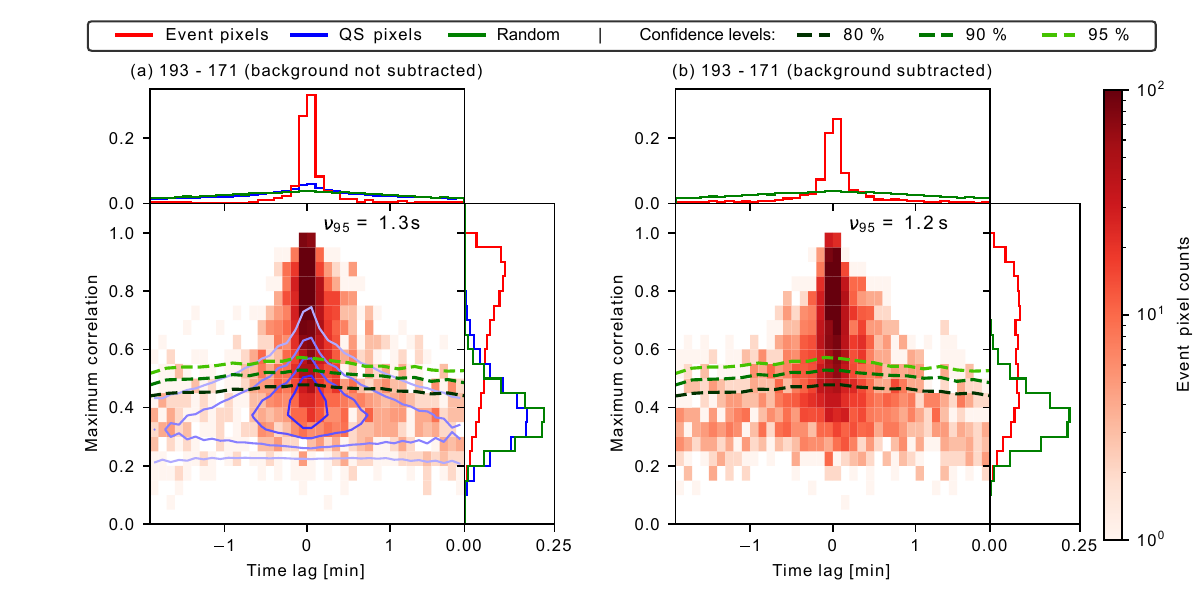}
    \caption{Example of the time-lag analisys results which show the margin and 2D histograms of time lags and associated maximum correlation values for the  193 \,–171 bands. Panel  {\it (a)} in {\it red} shows the
original distribution for the pixels involving the EUV brightening. Panel {\it (b)} shows the background subtracted brightening pixels. The {\it blue} contours in the central part of
subfigure {\it (a)} are the 20, 40, 60, and 80 percentiles of the QS pixel distribution. The {\it green} colors in the main panels are the confidence levels, and
the distributions of the light curves used to compute them are shown with the same color in the margin histograms. The margin histograms were normalized by the total number of pixels. Reproduced with permission from \citealp{Dolliou23}, copyright by the author(s).}
\label{fig:Dolliou23_7}   
\end{figure*}

One of the key questions that has been pursued in the past few years is the estimate of the temperature reached by these features. This has been achieved without ambiguity using spectroscopic information, as presented in the next section. The multichannel information of  AIA and HRI\textsubscript {EUV} can also be used for statistical purposes, even though, being the bands non-monochromatic, with more uncertainties on the results. 
 \cite{Dolliou23}, for instance, have investigated the light curves  from different AIA channels using the $time-lag$ method, which provides the correlation for the delay between light curves from a pair of bands sensitive to different temperatures.  
 They showed that the timescales for the evolution of these features are within the temporal resolution of the instrument (12 \,seconds) and that the surrounding QS has a clear, distinct timescale of changes, as shown in Figure \ref{fig:Dolliou23_7}. This work made a careful examination of the possible effects of the QS emission in the statistics of the events, showing that it does not change the results (right panel in the figure). 
 The results from the combination of all of the coronal channels of  AIA have demonstrated the same result, which is a cotemporality of the light curves' behavior for most of the cases. 
Following known results from numerical simulations {\cite[e.g.][]{Viall_2015} in coronal loops},  this has been interpreted as possible evidence of transition--region emission, contrary to the expectation of coronal values given the peak sensitivity of the HRI\textsubscript {EUV} channel  (light curves from coronal regions should show evident delays in the peaks). To be clarified here that the transition\,--\,region is the region where the thermal conduction acts as an heating term in loops, contrary to the coronal part where it acts as cooling term (\cite{Viall_2015} and references therein). In the cooler quiet Sun we expect the transition \,--\,region to coincide with the region below $\approx 1 \mathrm{MK}$ (see also \cite{Dolliou25} for further discussion on this aspect). 
However, the width of the distribution  shown in Figure \ref{fig:Dolliou23_7} also supports that a minority of events with clear signatures of delays among the channels in both directions, which evidence for cooling or heating up to coronal values.  Thus, this confirms the existence of a minority of events reaching coronal temperatures.

The combination of all the results reported above leads to the  picture where these tiny brightenings are mostly signatures of impulsive dynamics at transition \,--\,region heights. In the next section, we will discuss the detailed analysis of a few events, mostly based on the anlysis of spectroscopic data providing further properties.

\subsection{Results on Selected Families of Events}
\label{sec:selected}

\begin{figure*}
    \centering
    \includegraphics[width=\textwidth,clip=]{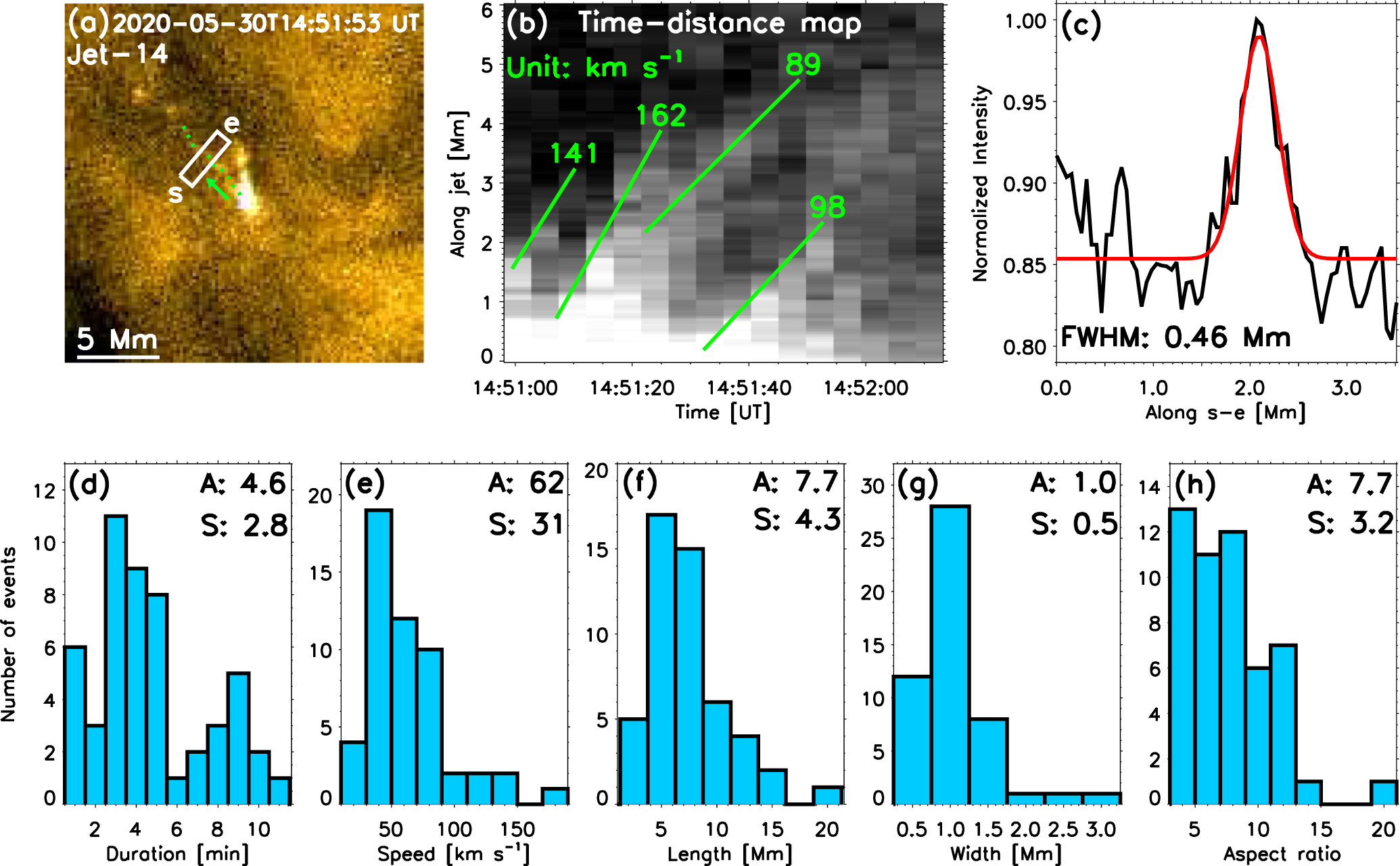}
    \caption{Result of a statistical study of microjets. (a) An HRI\textsubscript {EUV} image showing a jet in May 30 2020. The {\it green arrow} indicates the propagation direction, and the green dotted line marks the trajectory of the ejected plasma. (b) The time–distance map for the {\it green dotted line} in (a). Four tracks of ejected plasma and the corresponding projected speeds in the units of $\mathrm{km ~s^{-1}}$ are marked by the green lines and the numbers, respectively. (c) The intensity variation along the long side of the rectangle s\,--\,e shown in (a). The original intensity profile and the Gaussian fit are indicated by the {\it black and red curves}, respectively. (d)\,--\,(h) Distributions of the parameters for the microjets. In each panel, “A” and “S” represent the average value and standard deviation, respectively. Reproduced with permission from \cite{Hou21}, copyright by the author(s). }
\label{fig:Hou21} 
\end{figure*}

It is worth mentioning some works that made a selection criteria among the HRI\textsubscript {EUV} events automatically detected. This is the case for \cite{Hou21} who concentrated on microjets,  Y-shaped impulsive brightenings also observed over larger scales. They are often characterized by the expulsion of flow or blobs from the central spine. 
In their comparisons with the general EUV brightening population, they found their 52 features to be in the highest group for the size (7.7 $\pm$ 4.3 $\mathrm{Mm}$) and duration (4.6 minutes), as shown in Figure \ref{fig:Hou21}. Interestingly, they also derived the temporal evolution of the density and the emission measure-weighted temperature, the latter resulting to be  $ logT = 6.2$. This suggests that such events are part of the hotter distribution of the EUV brightening analyzed by \cite{Dolliou23}. 
From the derived plane of the sky velocity, density, size, and temperature, they inferred an energy budget of about $10^{24}$ $\mathrm{ergs}$, thus into the already sampled energy range.  
 These general properties where also confirmed by \cite{shi2024}, when 23 small jets were selected  manually and analyzed. Additionally, they provided the jets frequency distribution for thermal ($\approx 10^{23}$ \,--\, $10^{25}~ \mathrm{erg}$) and kinetic energies ($\approx 10^{21}$ -- $10^{24} ~\mathrm{erg}$), showing a negative power-law  behavior.

\begin{figure*}
    \centering
    \includegraphics[width=\textwidth,clip=]{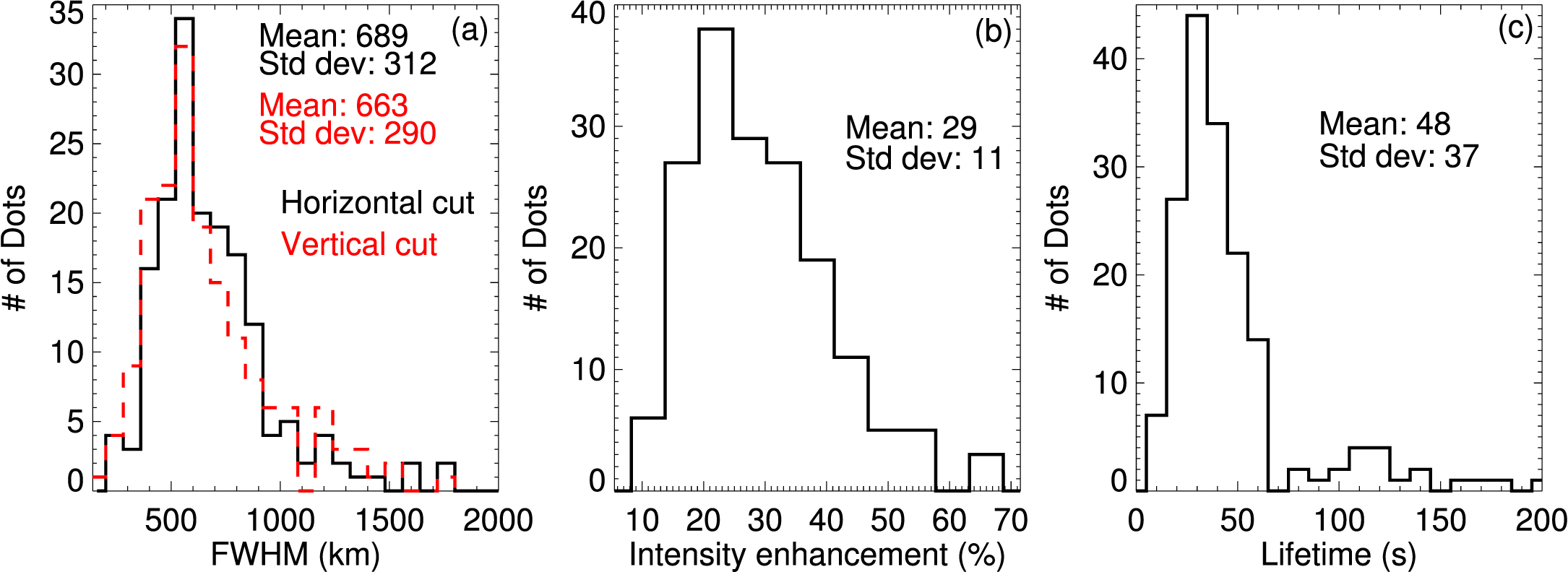}
    \caption{Histograms of the sizes, intensity enhancements, and lifetimes of 170 dots observed with HRI\textsubscript {EUV}. Panel (a) shows histograms of the horizontal and vertical sizes of dots. Panel (b) shows the histogram of intensity enhancements with respect to the immediate surroundings of dots. Panel (c) displays the histogram of lifetimes of dots. Reproduced with permission from \cite{Tiwari22}, copyright by the author(s).}
\label{fig:Tiwari22-3}   
\end{figure*}

Another example is from \cite{Tiwari22} who characterized the properties of 170 HRI\textsubscript {EUV} dot-like events observed by HRI\textsubscript {EUV} in emerging--flux regions within coronal bright points (see also Figure \ref{fig:tiwari22}). Figure \ref{fig:Tiwari22-3} shows the results of their statistical work, providing a mean size 675 $\pm \,300 ~\mathrm{km}$, and a lifetime of 50  $\pm \, 35$ seconds, thus well within the  properties of the whole HRI\textsubscript {EUV} brightening population. 
Half of these events were associated with larger loop brightening or larger jets. 

\begin{figure}
    \centering
    \includegraphics[width=0.5\textwidth,clip=]{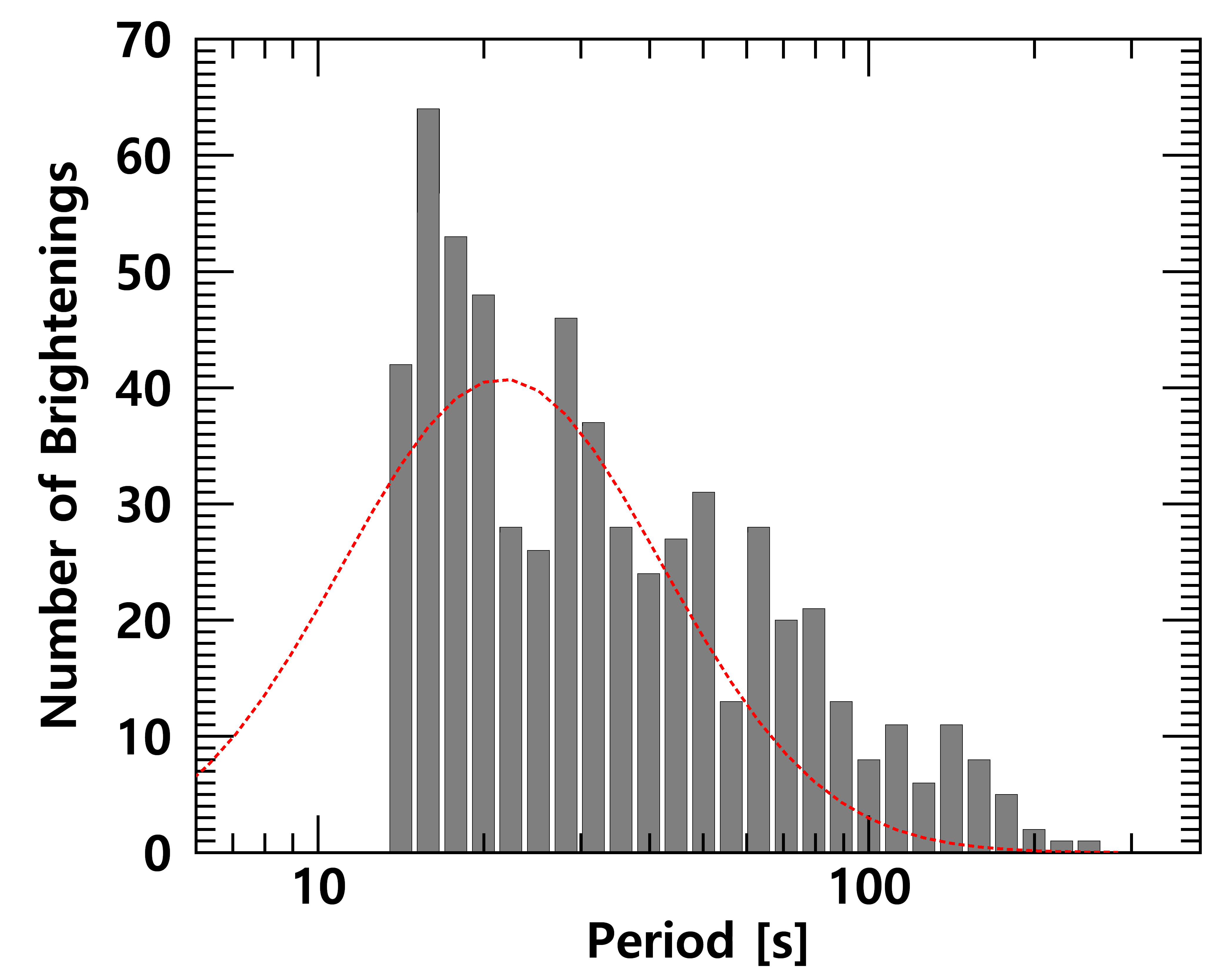}
    \caption{Logarithmic histogram of the  QPP periods in the HRI\textsubscript {EUV} brighenings. The lower limit of 15 seconds is given by the imposed cadence of the data.     
    The {\it dashed red line} represents the normal distribution in log space, with a mean of 21.6 seconds for QPP periods observed in standard solar flares \citep{Hryes2020}. Reproduced with permission from \cite{Lim2025}, copyright by the author(s).}
\label{fig:Lim25}   
\end{figure}

\cite{Lim2025} studied  the  possible presence of quasi-periodic oscillations (QPPs), usually detected in solar and stellar flares. This is a property of the emission that varies in a periodic or quasi-periodic way over a wide range of wavelengths. Their origin is still not completely understood, but the three main candidates are MHD waves, successive magnetic reconnections, and periodic reconnections triggered by waves. These authors used the \cite{Narang_2025} dataset and 
found that about 2.7$\%$ of the brightenings presented such a property. They also demonstrated that their  rate of occurrence increases (up to about 20$\%$) with the increase of the total event area, lifetime, and peak brighteness.
The measured periods are distributed within the 15\,--\,255 seconds range, with peak around 30 seconds, as shown in Figure \ref{fig:Lim25}. These appear to be within the values found for much larger flares. Similarly to larger flares, they found no correlation between period and peak  brighteness. Furthermore, contrary to  larger flares, they found  no correlation of the period with lifetime.  Given the latter results, the authors exclude standing waves (a candidate to initiate QPPs in flares) as a possible driver for QPPs.

\section{Plasma Properties from Individual Events}

The observation of these brightenings has evolved since their first detection by \cite{Berghmans2021}. Now there exist more complete datasets that include spectroscopic observations, from which we can infer more plasma properties. However, given the highly ephemeral and small-scale nature of these brightenings, combined with the generally limited field of view of the high cadence observations of spectrometers, the detection of these features is quite difficult and rare. 

\subsection{Temperature}
\label{sec:T}

\begin{figure*}
    \centering
    \includegraphics[width=\textwidth,clip=]{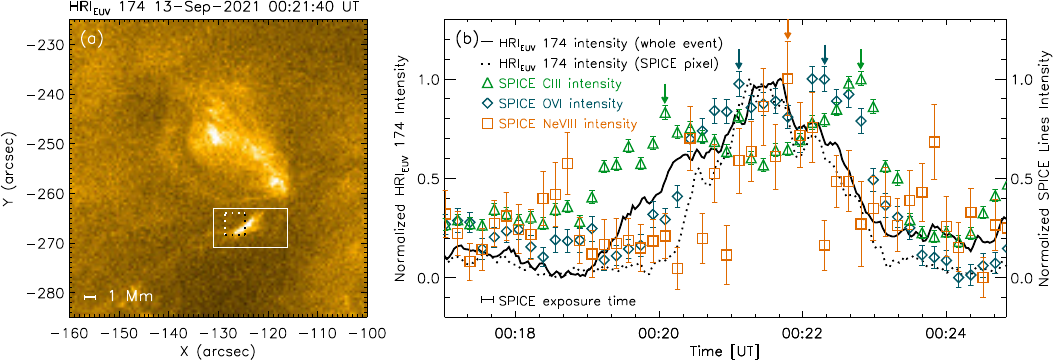}
    \caption{Normalized  SPICE light curves ({\it right}) of the event observed by HRI\textsubscript {EUV} on the {\it left} panel. Their peaks are marked by arrows in different colors. Reproduced with permission from  \cite{Ziwen23}, copyright by the author(s).}
\label{fig:ziwen23}   
\end{figure*}

\begin{figure*}
    \centering
    \includegraphics[width=\textwidth,clip=]{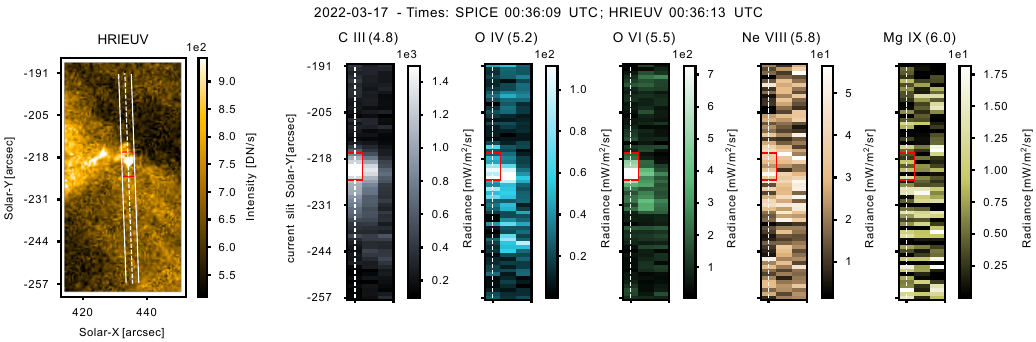}
    \caption{EUV brightening detected by both  HRI\textsubscript {EUV} ({\it left panel}) and the {\it SPICE}. The position of the SPICE slit is displayed as a dashed line on the HRIEUV and the SPICE images. The two white lines in HRI\textsubscript {EUV} delimit the field of view of the SPICE scan. The {\it red rectangle} is the event region. The temperature (log$\mathrm{T}$) of the maximum emissivity of each SPICE line is indicated within parentheses above each image. The latitude on the y-axis of the SPICE images refers to the position of the slit marked with a {\it white dashed line}. Reproduced with permission from  \cite{Dolliou24},  copyright by the author(s).}
\label{fig:Dolliou24-2}   
\end{figure*}

Few works  now exist where few HRI\textsubscript {EUV} brightenings have been observed also on SPICE, the spectrometer on board the  Solar Orbiter. Figure \ref{fig:ziwen23} from \cite{Ziwen23} shows the event detected on HRI\textsubscript {EUV} (left), and the light curves  from several spectral lines emitted at different temperatures (in the range  $4.8< \log T<5.8$) as measured by SPICE. These are compared to those from HRI\textsubscript {EUV}. Surprisingly, the HRI\textsubscript {EUV} event has response in all of the cooler transition \,--\,region  lines, while it is barely visible in the hottest Ne {\sc VIII} line, with a temperature of formation close to that of the peak of sensitivity of the HRI\textsubscript {EUV} band. This result has been later found also in most of the other events, as shown, for instance, in Figure \ref{fig:Dolliou24-2}, which shows no brightening signature in Ne {\sc VIII} and the hotter Mg {\sc IX} lines. These are examples that confirm the 
suggestion by \cite{Dolliou23}, that most of the HRI\textsubscript {EUV} events do not reach coronal temperatures, and what is seen in this band comes from the cooler lines falling within it.

\begin{figure}
    \centering
    \includegraphics[width=0.7\textwidth,clip=]{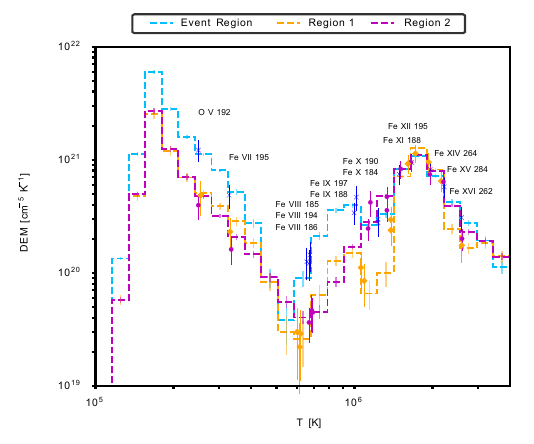}
    \caption{Differential emission measure as function of temperature for an EUV brightening deduced from  Hinode/EIS spectroscopic data. Reproduced with permission from \cite{Dolliou24},  copyright by the author(s).}
\label{fig:dolliou8}   
\end{figure}

The spectroscopic analysis of \cite{Dolliou24} over nine events has confirmed such a finding: combining SPICE data (mostly sensitive to transition\,--\,region emission) and Hinode/EIS data mostly sensitive to coronal emission, it was found no emission above about 1 $\mathrm{MK}$. This is shown in Figure \ref{fig:dolliou8}, where the differential emission measure for the brightening ("event region") has no emission above such a temperature. 
Furthermore, the light curve analysis from different lines showed no time delay among the light-curves, even though various general different behaviors were detected. In some cases, multiple brightening in the same area were seen, where all the light-curves followed the same path. In other cases, a cool peak was followed by a hotter one. 
These different behaviors were also reported by \cite{Nelson23}, where  15  HRI\textsubscript {EUV} events were observed using  the  IRIS spectrometer, which provides high resolution spectroscopy of the chromosphere and transition\,--\,region. Their Table 1 lists the main  properties derived, showing how only part of the HRI\textsubscript {EUV} events have chromospheric and/or transition\,--\,region counterparts. 

 The presence of hot and cool EUV brightenings was also found in coronal bright points investigated by \cite{Tiwari22}. They derived the differential emission measure using AIA data and found that generally hotter dot-like events were the largest in their dataset, reaching 1\,--\,2 $\mathrm{MK}$ in temperature. Thus, it is possible that the cooler population of the HRI\textsubscript {EUV} events happens in the most quiescent regions of the QS. 
Thermal analysis of small jets instead provides mostly plasma around or above 1 $\mathrm{MK}$ \citep{shi2024, Hou21}.

As we can conclude from these reported results, there is still need for further investigation to better characterize the thermal behavior of the different small--scale burst phenomena of the QS.

\subsection{Velocity}

\begin{figure*}
    \centering
\includegraphics[width=0.107\textwidth,clip=]{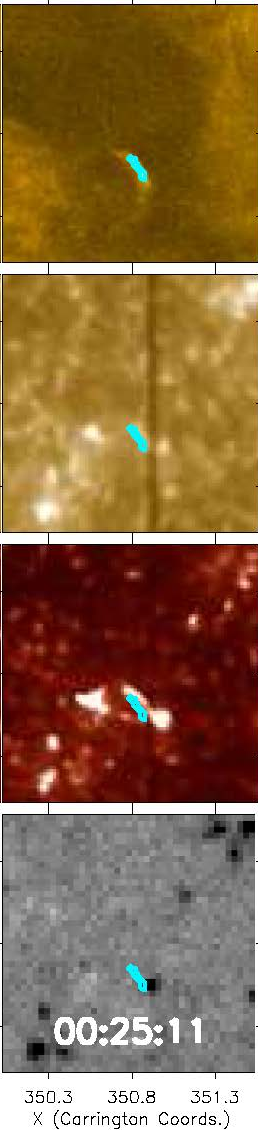}
    \includegraphics[width=.7\textwidth,clip=]{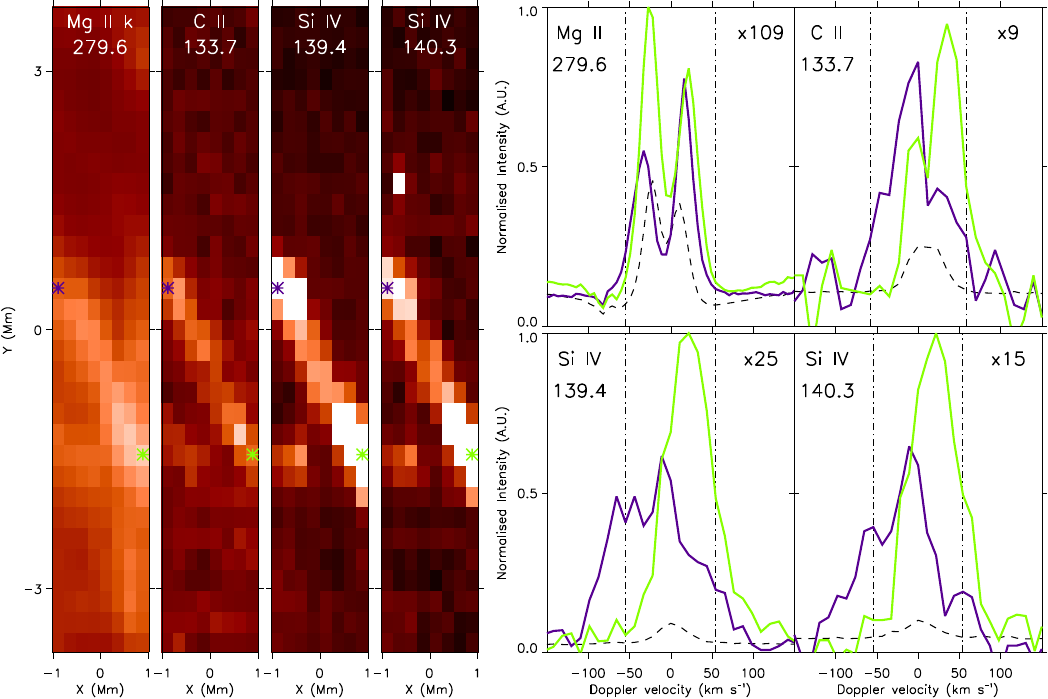}
    \caption{Spectral response to the extended EUV brightening  ({\it left column}). The {\it middle four panels} display the integrated
intensity over 110 $\mathrm{km s^{-1}}$  Doppler velocity windows around the rest wavelength for the Mg {\sc II} 279.6 nm ({\it first panel}), C {\sc II} 133.7 nm ({\it second panel}),
Si {\sc IV} 139.4 nm ({\it third panel}), and Si {\sc iv} 140.3 nm ({\it fourth panel}) spectral lines. The four panels on the right display the spectra sampled at the positions
of the {\it coloured crosses in the left panels}  (within the first and last raster positions). The {\it dashed black lines} displays the average spectra calculated from
this raster, whilst the two vertical {\it dot-dashed lines} book-end the Doppler velocity windows integrated across to construct the left panels. Each
spectral profile has been normalized. Reproduced with permission from \citep{Nelson23},  copyright by the author(s).}
\label{fig:nelson23}   
\end{figure*}

 \cite{Nelson23} reports the first Doppler analysis of the HRI\textsubscript {EUV} brightening, as illustrated in Figure \ref{fig:nelson23} using the  IRIS Si IV 1403 \AA ~line.  Doppler values between $\pm 20~  \mathrm{km~ s^{-1}}$  in both directions were measured, and about $\pm 100~ \mathrm{km~ s^{-1}}$  of Doppler width.   Knowing the thermal and temporal behavior of the event, the authors were able to classify one brightening as an explosive event (EE), already known in literature.

 Several works derive dynamic properties of the HRI\textsubscript {EUV} events from the time-distance plot of their intensity, which allows to infer the projected speed in the plane of the sky.
 \cite{Duan25}, using 59 detected HRI\textsubscript {EUV} events in the $\approx$ 2\,--\,8 $\mathrm{Mm}$ range,  showed upward-propagating disturbances with and  average velocity  of 62 $\mathrm{km~s^{-1}}$. 
 They found that most of the properties were similar to network jets. Indeed, as we can see from Figure \ref{fig:Hou21}, the tiny jets investigated by \cite{Hou21} have projected speed distribution on similar values. 
The same technique was adopted by \cite{Tiwari22} to analyze the dynamic of the dots within  bright points (see also Section \ref{sec:gen_br}). They revealed either small random proper motion or propagating motions, with values inferred in the plane of the sky of about 2\,--\,10 $\mathrm{km~s^{-1}}$   and up to  30 $\mathrm{km~s^{-1}}$, respectively.



\section{Link to the Photospheric Magnetic Field}

In order to build a complete picture of the origin and the whole evolution of these brightenings, we want to understand if there exists a connection with the underlying layers and if this is their trigger region. A candidate process for this impulsive emission is the changing of the photopsheric magnetic field due to emergence, cancellation, or simply magnetic-field reconfiguration as a consequence of surface flows. A review of a  few advances on this topic from  Solar Orbiter  follows.

There is an extensive bibliography on the investigation of the magnetic origin of known EUV brightenings, such as presented in the review by \cite{Young2018} for the chromospheric and transition \,--\, region events. Magnetic reconnection seems to be the main cause, from the "U" (Ellerman (EB), and IRIS bombs) or "$\Omega$" shaped configurations, where the two photospheric polarities converge. Reconnections at the spine from a fan configuration were also detected \citep[e.g.][]{Peter2019}. Magnetoacoustic waves are also invoqued as a disturbance due to the interaction of  magnetic-flux emergence or cancellation with the already existing magnetic field \citep[e.g.][]{Martinez15}.

Given the short temporal and small spatial scales involved for these measurements, 
the capability of the instrument to follow the changes of the magnetic field is a key factor in clearly identifying the origin of these brightenings. 
This is highlighted by, for instance, \cite{Chitta2023} who compared high resolution photospheric magnetic field variation (using PHI/HTR, $\approx~210 ~\mathrm{km}$), in the location of small QS loops observed by HRI\textsubscript {EUV}. They concluded that the heating and varying emission pattern of the HRI\textsubscript {EUV} emission ($\approx$ 30 seconds timescale) are probably associated to the weak field, small-scale patches of mixed polarity, most of the time found at or nearly loops footpoints. They also localized such weak patches to be dominant at the edge of stronger field areas (see their Figure 1).

There are different kinds of approaches proposed in literature to investigate the link between the photosphere and the upper atmosphere, and the resulting picture is still not  conclusive. 
Some of the works select  EUV events based on their morphology or typology, and then study the corresponding photospheric area. Others group the EUV events based on the similarities of their photospheric properties. 

\cite{Hou21}, mentioned in Section \ref{stat}, made their HRI\textsubscript {EUV} statistical analysis of microjets using  SDO/HMI \citep{Scherrer2012} data and concluded that  such events could be found both in clear bipolar regions, and regions where an inclusion of a minor polarity was seen within a larger scale unipolar region. 
 Similarly to \cite{Berghmans2021} and \cite{Zhukov2021}, the network lane was  the main location where these EUV brightenings are found,  or at least one of the footpoints of the associated magnetic structure lies within this region. \cite{Zhukov2021} identified few events having very weak magnetic field, below 10 $\mathrm{G}$.

Similar magnetic properties were found by 
\cite{Panesar21} who manually  identified 52 EUV brightenings using both HRI\textsubscript {EUV} and AIA. These were of various shapes, not just microjets.  
Analyzing the  HMI temporal evolution below these events, they could affirm that most of them were located in  network lanes, above neutral lines, often ($79\%$) overlaid by a cool mini-filament. Most of them manifest magnetic-flux cancellation at a rate of $\approx 10^{18}~~\mathrm{Mx ~hr^{-1}}$.

\begin{figure}
    \centering
    \includegraphics[width=0.7\textwidth, clip=]{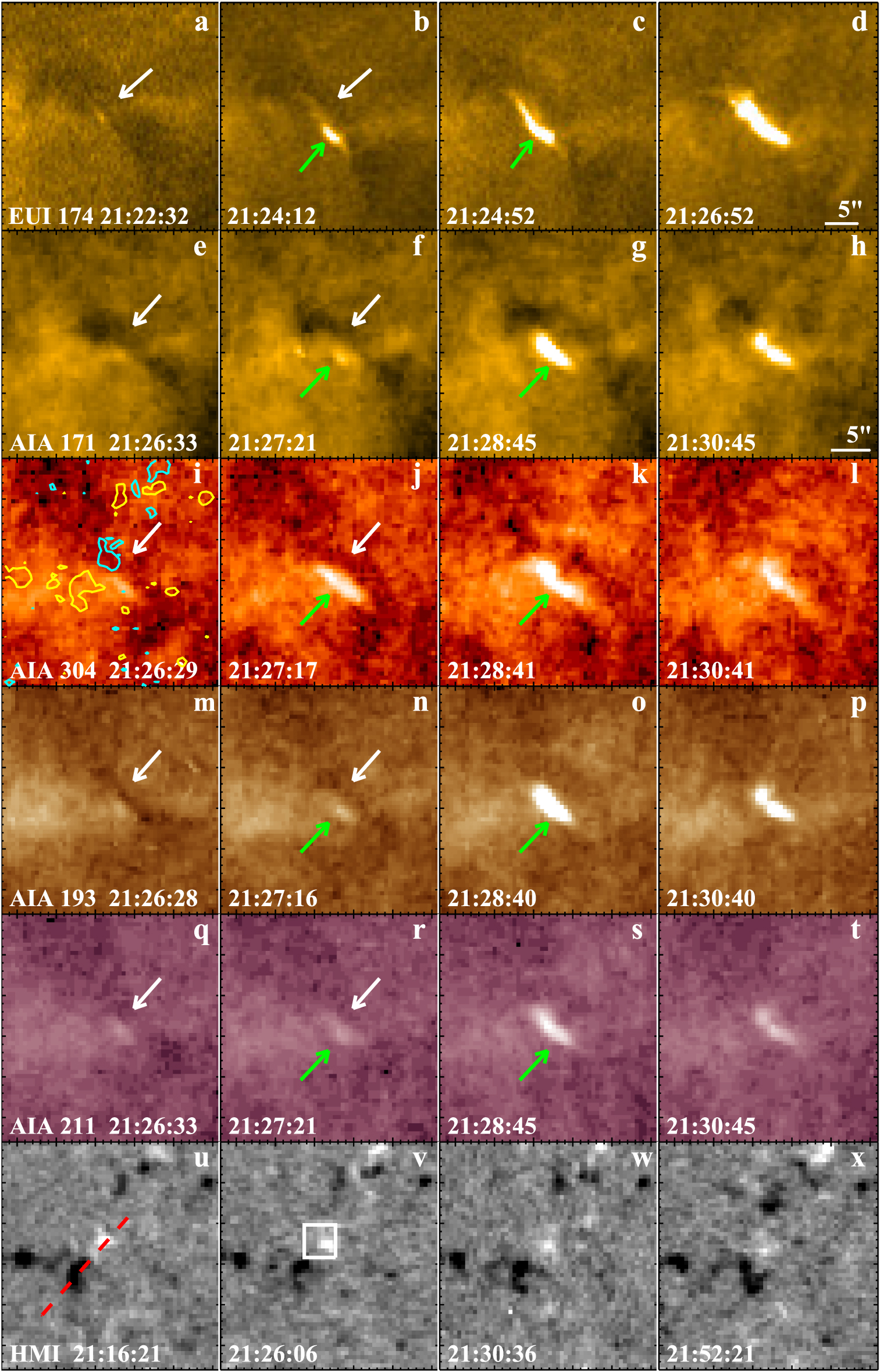}
    \caption{Example of the temporal evolution of a brightening on  May 20 2020, observed in the HRI\textsubscript {EUV} and AIA channels ((e)\,--\,(t)), together with the line-of-sight magnetic field friom HMI. Panels (a)\,-–\,(d) show 174 Å HRI\textsubscript {EUV} images of the campfire and display the same field of view as inside the {\it white box } of Figure 1(c).  The {\it white arrows} point to the cool-plasma structure, and the {\it green arrows} point to the brightening itself. In panel (i), HMI contours, of levels ±15 G (at 21:26:06) are overlaid, where {\it cyan and yellow contours } represent the positive and negative magnetic polarities, respectively.  Reproduced with permission from \cite{Panesar21},  copyright by the author(s).}
\label{fig:panesar21}   
\end{figure}

Among the group of works that make a classification based on the properties of the photosphere,   \cite{Kahil2022} used  the high-resolution instrument  PHI on  Solar Orbiter (which provides a spatial resolution of about 200 km in this case). They revealed a more complex scenario of the HRI\textsubscript {EUV} events, proposing two  groups from their 38 detected events. 
The first one collected the events that were seen in  projection above a clear bipolar region with signatures of magnetic reconnection; these are the majority of the population. They suggested similarities to EB EUV bursts.  
The second group collected all the rest of the events where the photospheric connection was not clearly identifiable. The latter include either the lack of magnetic polarities (probably due to a too weak field) or the presence of randomly scattered magnetic polarities, showing  dynamics resembling convergence, emergence, appearance, and disappearance. They suggested that possible reconnection could arise higher in the atmosphere than for the first category. 
Importantly, they could not identify a clear correlation between the HRI\textsubscript {EUV} events' shape and the photospheric configuration, in agreement with other works. 

\begin{figure}
    \centering
    \includegraphics[width=\textwidth,clip=]{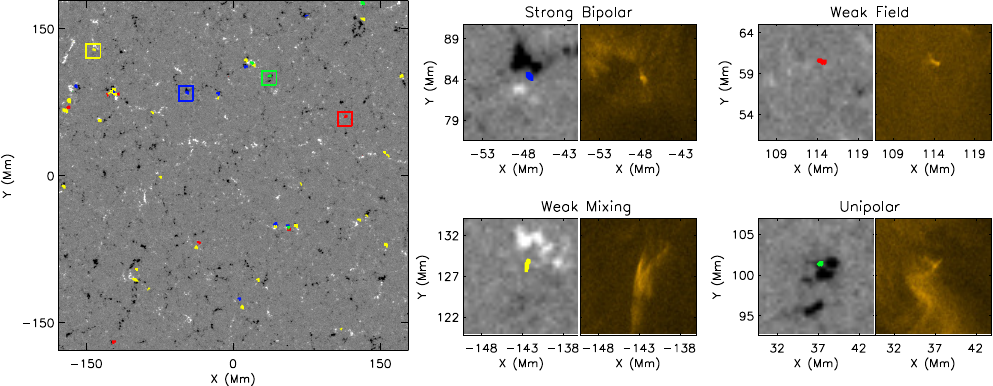}
    \caption{Examples of the different magnetic field properties below the EUV brightenings. {\it Left panel}: Background image of a line-of-sight magnetic field map measured by SDO/HMI, saturated at ±100 G. The overlaid contours outline EUV brightenings with the different colors shown on the right. From \citep{Nelson24a},  copyright by the author(s).}
\label{fig:nelson24}   
\end{figure}

\cite{Nelson24a} carried out a statistical analysis comparing HRI\textsubscript {EUV} events to SDO/HMI line-of-sight magnetic field of about 5000 events. They grouped these based on the strength of the  line of sight magnetic field  and found that, contrary to \cite{Kahil2022}, those associated with bipolar regions were in the minority, and only a small part of them showed emergence or cancellation. 
The larger number of events were those having the strongest absolute magnetic strength (20 G) close to the photospheric network. Figure \ref{fig:nelson24} shows the variety of magnetic field configurations below the HRI\textsubscript {EUV} events.  An interesting result of this work is the similarity of statistical behavior, for most of the selected groups, of the distributions for the projected area, total intensity and duration for both the HRI\textsubscript {EUV} events and the associated magnetic field. This is an encouraging result on the method to associate phenomena between the atmospheric layers.

%

%

 \section{Origin of EUV Brightenings}
\label{sec:gen_br}

 Few theoretical works address the origin of these features, trying to identify the physical processes behind them. The approach can be statistical on the whole automatically detected events, or focused on specific subgroups.

The statistical approach was used by \cite{Barczynski2022} who
simulated the formation of small scales energy dissipation regions in QS, by  perturbing the extrapolated photospheric magnetic field subject to energy dissipation by the magnetofrictional process. 
This work provides the statistical properties of the dissipated energy along the line of sight over detected concentration areas, and compares them to what obtained by observing the HRI\textsubscript {EUV} brightenings in a QS. 
Their probability distribution function for projected area, major axis, total intensity, volume and other quantities (see their Figure 4), are very similar to those derived from their observations (and those reported in Figure \ref{fig:Narang25}, for instance). 
Beside the smaller amount of detected events in the observations (240) with respect to what find in other observational works presented in Section \ref{sec:pdf}, the cited quantities resemble the known power-law  distribution. The simulated quantities follow the same distributions,  with the major difference being a larger (19 -- 29 times) number density detected in the simulations. This has been attributed to the simple physical approach of the model, where mostly the energy is dissipated through the frictional processes, with no energy losses trough radiation or conduction. However, the general statistical properties found are encouraging, and further investigations are needed with a more complete physical approach.

There exist several works which address the origin of specific small scale features among those collected by the whole group of HRI\textsubscript {EUV} brightenings. 
In this regard, \cite{chen21} and \cite{chen2025}   use the same setting of the 3D MHD MURaM model, which mimics the small-scale dynamic evolution of the quiet Sun and synthesizes the emission from the HRI\textsubscript {EUV} band. 
\cite{chen21} showed that the model was able to produce various types of transient brightenings resembling those of HRI\textsubscript {EUV}. In particular,  some have signatures of localized small\,--\,scale heating, where the plasma reached 1\,--\,2 MK. Their emission was localized in the eights of 1\,--\,4 Mm range. They classified seven brightening in their simulation in three groups given by the magnetic configuration (highly twisted flux rope, crossing bundles, forking bundles). Their interpretation on the origin of the brightening was a reconnection happening at coronal level, rather than down at the bundle footpoints as it can arise, for instance, for Ellerman bombs \citep{Young2018}. Furthermore, they did not find evidence for magnetic field emergence  or cancellation, evoked instead in other works (see later). 
 \cite{chen2025} instead concentrated on those brightenings having plasma properties similar to explosive events (bidirectional reconnection flows). Their investigation, which included simulated spectroscopic observations, concluded that  there can be both hot and cool events, depending on the local magnetic-field conversion to heat or to the local density.

\cite{Amari2025} investigated 
the possible role of instability and eruption of small scale twisted flux ropes in QS to explain the existence of these EUV brightenings. In their extrapolation with XTRAPOL code, they demonstrate that small twisted flux ropes are created easily in a force-free magnetic extrapolation from photospheric measured magnetifield. 
The estimated height and stored free energy are consistent with those derived from the HRI\textsubscript {EUV} data. Furthermore, they found conditions (number of twists, magnetic stored energy) which are known to lead to instability and eruptions in large scale flux ropes, such as coronal mass ejection. Using an MHD high resolution simulations (30  $\mathrm{km}$), they also demonstrated that small scales dynamic flux ropes form by the presence of a local dynamo activity below the photosphere, with flux emergence, flux cancellation and the rope eruption. 

The brightening were instead interpreted as coming from short (within the  observed heights from HRI\textsubscript {EUV}) loops by \cite{Dolliou25}, who investigated them using the HYDRAD 1D code.
They analyzed the thermal response and synthetic brightness' temporal behavior from imagers and spectrometers. They demonstrated that the observed co-temporal peaks of the brightness from emission at different temperatures come mostly from short-loops being impulsively heated from an initial cool equilibrium state. When the heating arises in short loops that are already hot, this observational signature is only reproduced for strong heating, producing also strong plasma flow. This case  reproduces signatures found in observed explosive events, similarly to \cite{chen2025}.

A  Bifrost magnetohydrodynamic (MHD) simulation was used to investigate two groups of small scale events: dot-like and tiny jets. These works included the calculation of the synthetic Fe IX-X intensities, which are found at the peak of the HRI\textsubscript {EUV} response function and the analysis of HRI\textsubscript {EUV}, SDO/AIA and HMI. 
From this simulation, \cite{Tiwari22} showed that the tiny, bright dot-like events can be formed during the emergence of flux in the quiet Sun. They were able to reproduce several properties: the location near strong magnetic flux patches or near neutral lines, their temperature (hot and cool), size, and their plasma flow. They concluded that both reconnection and magnetoacoustic shocks were compatible with their data, as the origin of those events. 

The small scale coronal jets in quiet Sun network regions were instead studied by \cite{Panesar23}.  They identified five jets having almost all similar qualitative properties of those observed: morphology, size at the base and spine length, duration, speed at the spine,  and presence of cool plasma. Observing the temporal behavior in the simulation, they concluded that the magnetic convergence and cancellation at the neutral line are responsible (four out of five jets) of the jet triggering, similarly to coronal jets. Reconnection-driven process is evoked to explain such findings, also supported by the synthetic  Doppler blue and red shifts at the spine site (suggesting untwisting motion). 
The hotter temperature from the simulation was reaching coronal values, while the cool plasma evolution had mini-filament eruption behavior. Thus, similarly to what deduced by \cite{Hou21} (Section \ref{sec:pdf}), these resemble  the small version of coronal jets.

In summary, there is an undergoing effort to reproduce the observed EUV brightenings through numerical simulations and MHD modeling, with results that are not exclusive. This aspect has to be taken into account and improved, for allowing future progress in this topic.

\section{Conclusions}

This review reports the collection of new observational  and theoretical results from Solar Orbiter instruments on the smallest scale EUV brightenings identified by the instrument  EUI/HRI\textsubscript {EUV} in the QS. 
Taking the $\approx $ 0.4\,--\,5 Mm ejected spatial scales, we find a forest of various events of different shapes (microjets, loops, dots, or more complex configuration) and duration (3 seconds - minutes),the latter  being limited by the available cadence and length of the temporal sequences analyzed. 
Quantities such as size, area, lifetime, brightness  kinetic and thermal energies (at least for the microjets) follow a negative power- law probability distribution, resembling the continuation at small scales of the well-known flares-microflares EUV and X-ray distributions in the corona. 
These provide evidence of the self-similarity  of impulsive brightenings that provides an interesting insight to be further investigated.

The thermal analysis revealed that, even if all of the events are detected by the same HRI\textsubscript {EUV} channel, their thermal character may be different. The majority of them are probably cooler than 1 MK, thus questioning their role in  coronal heating. Their rate of appearance  is also not high enough to compensate for the radiative losses of the corona. 
Their low temperature  would be in agreement with the deduced height above the photosphere, which is within 5  Mm (recall that this  value is a consequence of the parameters imposed in the event selection criteria,  see Section \ref{sec:gen}).
Thus, if these tiny brightenings arise mostly in the chromosphere and transition\,--\,region,  the observed power\,--law s indicate that the self-similarity is a property of the whole solar atmosphere. 

Several of the  works mentioned have found resemblance with known impulsive brightenings, such as EB, microjets, flaring loops, and EUV bursts. Some cases are a smaller version of them.  
Again, this demonstrates that the physical processes involved are independent of the spatial and energy scales. This is something that would be interesting to further investigate. To do so, it would be useful to separate the detected events into sub-categories and investigate them independently  as done in some of the works cited in the previous sections. This would probably bring more order in the proposed data interpretation and understanding the physical processes involved. Numerical and MHD simulations have been able to reproduce several of these various  observed properties. These are mostly the results of magnetic reconnections, even though the height in the atmosphere (corona or lower down) where this happens may vary. It would be useful to continue the investigation and clarify these aspects, in order to understand if this is model dependent. Furthermore, photospheric magnetic emergence and convergence are often evoked, even if not in all the cases. This is also another open point. 
The production of more observational outputs from the simulations should help for these  opened questions.


As we can conclude from these few examples, which attempt to explain the origin of a few subgroups of HRI\textsubscript {EUV} events, there is still much work to do  that makes a connection with the already known small-scale bursts observed in the solar atmosphere. These new observations at very high temporal and spatial resolutions made by HRI\textsubscript {EUV} and PHI, coordinated with those from spectroscopic instruments, should be used to add the missing information: dynamics and magnetic changes at high cadence at small scales. We could elucidate the underlying, fundamental processes at work.  The role of the small scale photospheric magnetic field, some of which probably still unresolved, has an evident role in the energy transfer from this layer  to the upper solar atmosphere, thus its investigation is extremely important to understand its possible contribution to heat the corona. To be noted that the today regular observations by  PHI still measures a flux higher (3.5 $\times \mathrm{10^{15} ~ Mx}$, \citealp{Chitta2023}) than what provided by SUNRISE (9 $\times \mathrm{10^{14} ~ Mx}$, \citealp{Solanki2010, Anusha2017}).

Additionally, the statistical work should be continued in other regions of the solar atmosphere to see if there is any variations due to the local properties of the environment. We should also try to understand why we are observing  self-similarity of the solar atmosphere. What are these physical processes that dominate it and persist at all scales? 

\begin{acks}

The author acknowledges the anonymous referee for the constructive comments on the manuscripts. 
Solar Orbiter is a space mission of international collaboration between ESA and NASA, operated by ESA. The EUI instrument was built by CSL, IAS, MPS, MSSL/UCL, PMOD/WRC, ROB, LCF/IO with funding from the Belgian Federal Science Policy Office (BELSPO/PRODEX PEA 4000134088, 4000112292, 4000117262, and 400013447); the Centre National d’Etudes Spatiales (CNES); the UK Space Agency (UKSA); the Bundesministerium für Wirtschaft und Energie (BMWi) through the Deutsches Zentrum für Luft- und Raumfahrt (DLR); and the Swiss Space Office (SSO).
This work was supported by the International Space 1109 Science Institute (ISSI) in Bern, through ISSI International Team project \#23- 1110 586 (Novel Insights Into Bursts, Bombs, and Brightenings in the Solar Atmo1111 sphere from Solar Orbiter).
\end{acks}




\begin{ethics}
\begin{conflict}
The author declare that they have no competing interests.
\end{conflict}
\end{ethics}


\bibliographystyle{spr-mp-sola}
\bibliography{biblio}

\end{document}